\documentclass[sigconf]{acmart}
\acmConference[ESEC/FSE 2019]{The 27th ACM Joint European Software Engineering Conference and Symposium on the Foundations of Software Engineering}{26--30 August, 2019}{Tallinn, Estonia}

\usepackage[utf8]{inputenc}

\usepackage{algorithm}  
\usepackage{algorithmic} 

\title{NeuralVis: Visualizing and Interpreting \\Deep Learning Models}

\author{Xufan Zhang, Ziyue Yin, Yang Feng, Qingkai Shi, Jia Liu, Zhenyu Chen}
\affiliation{%
  \institution{State Key Laboratory for Novel Software Technology at Nanjing University}
  \city{Nanjing}
  \country{China}}
\email{liujia@nju.edu.cn,  zychen@nju.edu.cn}

\newcommand{\yang}[1]
{
   {\noindent\color{red}\bf [#1]$_{\scriptscriptstyle\textit{yang}}$}
}
\newcommand{\xufan}[1]
{
   {\noindent\color{blue}\bf [#1]$_{\scriptscriptstyle\textit{xufan}}$}
}

\newcommand{\includeAuthorComments}[1]
{
   \ifthenelse{\equal{#1}{0}}
   {
      \renewcommand{\yang}[1]
      {
         {} 
      }
      \renewcommand{\xufan}[1]
      {
         {} 
      }
   }{}
}

\begin{document}

\begin{abstract}
Deep Neural Network(DNN) techniques have been prevalent in software engineering.
They are employed to faciliatate various software engineering tasks and embedded into many software applications.
However, analyzing and understanding their behaviors is a difficult task for software engineers.
In this paper, to support software engineers in visualizing and interpreting deep learning models, we present NeuralVis, an instance-based visualization tool for DNN.
NeuralVis is designed for: 
1). visualizing the structure of DNN models, i.e., components, layers, as well as connections;
2). visualizing the data transformation process; 
3). integrating existing adversarial attack algorithms for test input generation;
4). comparing intermediate outputs of different instances to guide the  test input generation;

To demonstrate the effectiveness of NeuralVis, we conduct an user study involving ten participants on two classic DNN models, i.e., LeNet and VGG-12.
The result shows NeuralVis can assist developers in identifying the critical features that determines the prediction results.

Video: \url{https://youtu.be/hRxCovrOZFI}

\end{abstract}
\keywords{visualization, neural network, comparative research}

\maketitle

\section{Introduction}
Deep Neural Networks have been employed to develop many DNN-based applications because of its high accuracy and superior performance in handling some well-defined tasks, including natural language processing(NLP), image classification, and face recognition.
However, because DNN-based software applications are constructed based on the design and programming diagrams that are different from the conventional software systems, it becomes difficult for developers to analyze and further understand their behaviors and execution in the development.
Further, the lack of interpretability raises people's concerns about reliability of DNN-based applications. 
In consequence, it is difficult to put into use, especially in some safety-critical fields. 
To alleviate these problems, researchers have proposed many software testing and debugging methods to improve the model quality.
However, practices in deep learning testing are still in the early stage~\cite{pei2017deepxplore}, almost all of these approaches focus on finding adversarial examples based on the structure coverage~\cite{tian2018deeptest}\cite{zhang2018deeproad}\cite{ma2018deepgauge} but fail to provide interactive ways to support developers to guide the testing procedure. 
Especially, for safe-critical applications, assisting developers in understanding reasons why the model behaves correctly is a challenging but necessary task.

By transforming abstract data into graphics, visualization techniques is proved to be an extremely useful aid to study in fields like hydrodynamics\cite{potma2001real} or chaos theory\cite{highfield1996frontiers}. 
In deep learning, researchers have been working on approaches to visualize detailed information during the training stage. However, exploration in training stage focus on metrics to measure performance of representation learning other than the learned information.


Trained neural network models are saved in a complex file format. 
To bring software engineers intuitive understanding, detailed information is extracted, rearranged, transformed, and visualized. In addition, engineers can do few interactions with the model during the automatic training phase since hyper-parameters in the model keep tuning during each epoch. But interacting with the model is possible when the training is over. To assist engineers in interpreting a trained model, visualization is applied with the aim of revealing semantic features inside the model. Differences in the goal means that visualization for trained network model should be different from that for a model to be trained.

Based on the above motivations, we introduce NeuralVis -- an interactive visualization tool for trained neural networks. New approaches to investigate the trained neural network model are proposed with the guidance of comparative research. NeuralVis has three main features:
\begin{itemize}
\item[1)] We present an instance-based visualization approach for trained neural networks. This approach can visualize both the static structure and the dynamic behaviors of neural network models. 
\item[2)] We design interfaces to enable developers to interact with the visualized neural network. These interfaces facilitate developers in manipulating the model structure and thus analyzing their behaviors.

\item[3)] We implement our approach into a tool, namely NeuralVis, to validate our approaches. NeuralVis offers engineers interface to compare intermediate outputs of two inputs, to analyze activation status of a chosen layer, and to mutate the model by freezing filters in convolutional neural networks(CNN).
\end{itemize}

With our implementation of NeuralVis, We demonstrate its ability to visualize the structural architecture of the neural network. The algorithm to achieve model mutations by freezing filters is also explained in detail. Software engineers are able to conduct differential analysis via comparison on intermediate output and model mutation, the feedback proves the intuition brought to them. 

\section{Related Work}

\subsection{Visualization for the Model Structure}

For a trained neural network model, the architecture is a primary component to be visualized. Regarding the network structure visualization, the most common way is a node-link diagram, due to the popularity of TensorBoard and its interactive dataflow graph\cite{harley2015interactive}. However, the graph is not suitable for complex models as there will be numerous links if there is a large number of neurons in the network. TensorSpace\footnote{https://tensorspace.org} solves the problem via 3D visualization where neurons in each layer is distributed in space. 

The feature map is a popular component to be visualized in image classification tasks. The output value of a feature map is treated as a 2D array, which can be transformed into an image by nature\cite{zhang2018visual}. Image-resolution receptive field of neural activation is computed in \cite{bolei2015object} by inverting the feature map back to the input space. 

The logic for each prediction is another component researchers work on. The representation of CNN is disentangled into decision trees to quantitatively explain the logic for each prediction in \cite{zhang2018interpreting}. 

\subsection{Visualization for the Model Behavior}

TensorBoard displays the computational graph of the model, and it produces scalar values during the computation \cite{girija2016tensorflow}. 
NeuronBlocks empowers engineers to check model configuration and model architecture \cite{gong2019neuronblocks}. For these works, metrics like the loss, accuracy and training time are updated after every epoch. Whether the trained model is overfitting can also be evaluated with these metrics.
Facebook engineers and data scientists use visualization tools in their normal workflow \cite{kahng2018cti}. Visualization techniques are introduced to illustrate the function of intermediate feature layers and the operation of the classifier in \cite{zeiler2014visualizing}.



\section{Approach}
For trained network models, researchers have been working on visualization methods to make it understandable. Regardless of the high score it achieves over the whole dataset, how it predicts a single input is still hard to understand. To simplify the task, instance-based visualization is adopted. 

Several visualization tools are proposed to visualize deep neural networks, especially convolutional neural networks. We build our visualization tool based on the existing 3D visualization techniques provided by TensorSpace. Both 2D and 3D visualization are adopted in our tool. 3D visualization brings user an overview of the structural architecture for a trained model and 2D enables testers to concentrate on the detailed activation status on a specified layer. To make it more interactive, comparative research and differential analysis are introduced in our visualization tool. Engineers can compare intermediate output of different inputs to check the similarities and the differences, which adds to their comprehension of how adversarial samples deceives the neural network model.

NeuralVis is built on top of TensorSpace, thus we will not elaborate how model component is visualized in detail. Instead, we introduce how we bring more intuitions with our innovative approach. To bring engineers more intuition about trained model, we propose a novel approach to implement visualization by integrating the concept of comparative research. Comparative research is the act of comparing multiple things with a view to seek out any findings. In NeuralVis, two features to achieve comparative research in trained models are introduced.

\textbf{Feature 1: Output comparison}. 
For a given trained model $M$, let $x$ denote the input, $y$ denote the output. A common way to describe neural network is: $y = M(x)$.

Since Model $M$ consists of multiple layers, if we use $f_i$, $w_i$ and $b_i$ to describe the activation function, weight matrix, and bias of $i^{th}$ layer, it can be expanded as:
\begin{equation}\label{equ:expand}
y = f_n(w_n(f_{n-1}(w_{n-1}(..., f_1(w_1x, b_1), ...)), b_{n-1}), b_n) 
\end{equation}

Multiple filters in different layers can be frozen at the same time. The \textbf{mutate\_output} algorithm is shown in Algorithm \ref{alg:muou}. It takes as input the trained neural network $model$, the sample $input$, and the frozen filter configurations in JSON format $config$. Static structural information of layers are extracted from the model. Output of each layer is calculated in the loop, if no $config$ is present, the result will be passed to the next layer as input. Otherwise, function $prepare\_input$ is responsible for applying zero vectors on filters in that layer. Output of each layer is stored in a array that is returned once it reaches the last layer.

\begin{algorithm}[!h]
\caption{mutate\_output($model$, $input$, $config$)}
 \label{alg:muou}
\begin{algorithmic}[1]
    \STATE $structure=extract\_layers(model)$
    \FOR{$i=0$ to $len(structure)$}
        \STATE $output=inner\_output(model, input, i)$
        \IF{$config$ is $None$}
            \STATE $input=output$
        \ELSE
            \STATE $input=prepare\_input(output, config, structure, i)$
        \ENDIF
        \STATE $result[i]=output$
    \ENDFOR
    \RETURN $result$
\end{algorithmic}
\end{algorithm}

Intermediate output of layer $i$ can be calculated and visualized. At the moment, it is still difficult for engineers to tell how the network model works even with these intermediate outputs. For example, engineers can hardly tell why the model fails at a specific input. Rather than telling the causes of the prediction, by adding an additional input for comparison, engineers are expected to tell the differences between two inputs in intermediate layers. 

Common tasks like adversarial sample generation are also included in this tool since it is a useful aid to create test inputs. To learn more details about the different among different inputs, especially the differences between the original input and the adversarial sample, engineers can look into a specific layer to check the activation status.

\textbf{Feature 2: Model mutation}.
Model mutation is achieved by adding perturbations to equation \ref{equ:expand}. Filters in CNNs are considered to extract features from the original input. To remove the filter from the network, zero vector is applied on the weight. Let filter $f_{ij}$ denote the $j^{th}$ filter in layer $i$, $f_{ik}$ is the filter to be removed, let $output_{{i-1},j}$ denote the $j^{th}$ subvector of input of $i^{th}$ layer, $m$ denote the size of filters in $i^{th}$ then the output of $i^{th}$ layer is:
\begin{equation}
f_i(\sum_{j=0}^{k-1}w_{ij}output_{{i-1},j}+\sum_{j=k+1}^{m}w_{ij}output_{{i-1},j}, b_i)
\end{equation}

The filter is considered as frozen when a zero vector is applied on it.  Engineers can freeze filters in a convolutional neural network to mutate the trained network model, changes it brings to the following layers are visualized. These changes bring engineers intuition about the role a filter plays in the trained network. 

\section{Architecture}
\begin{figure}[h]
  \centering
  \includegraphics[width=\linewidth]{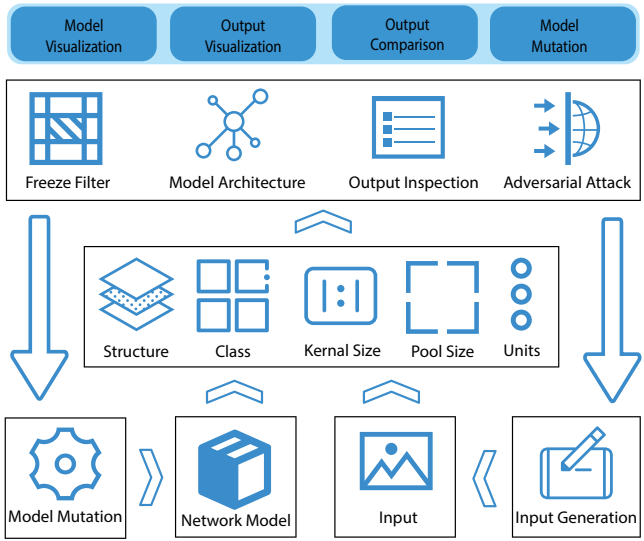}
  \caption{Components and features of NeuralVis}
  \label{fig:neuralvis}
\end{figure}

We present the architecture of NeuralVis in Figure \ref{fig:neuralvis}. Model visualization and output visualization are the fundamental functions. For a given trained model and input, static structural information is  extracted, how the input transforms inside the network is calculated. With these data, the model and the output are visualized. To make the process more interactive, NeuralVis reacts to engineers' operations in model mutation and input generation.

NeuralVis consists of three main sections as depicted in Figure \ref{fig:userstudy} where 1) the operation panel on the left to perform primary tasks, 2) the detailed panel to visualize details of the model structure and the data transformation throughout the model, 3) the comparison panel to intercept intermediate output of multiple inputs on a specified layer when the comparison task is activated. 

\subsection{The operation panel}
The operation panel shows primary tasks engineers can perform, including 1) model upload, 2) model selection, 3) input upload, 4) input selection, and 5) synthetic samples generation, 6) single input visualization, and 7) a pair of inputs comparison. As shown in Figure \ref{fig:neuralvis}, the network model and the input data are two primary components engineers work with. NeuralVis support Keras models at the moment. Uploaded network models appear in the model list. Trained network models are stored in hdf5 file which is not readable. NeuralVis calls the API in Keras library to load the model. Static structural information is extracted from the model, transformed into a 3D graph. However, only general configurations of the models are provided for a trained network models. NeuralVis handles tasks related to image classification, for simple images which can be created by hand easily, engineers can create stick figure consisting of lines and points with the provided sketchpad. 

In addition, engineer can leverage the sample generation function to create adversarial samples with particular attack algorithms, which is located at the bottom of this section. Available algorithms are provided in a option list. If an adversarial sample is generated successfully, it will be added to the sample list. Engineers waste much time in code implementation for adversarial samples. They are free from the implementation in coding in NeuralVis. All they need to do is operate on the panel to tell NeuralVis the original input together with the selected algorithm. 

\subsection{The detailed panel}
The detailed panel works as a response to user's actions in the operation panel. Detailed information about models and intermediate activation status will be presented to realize model visualization and output visualization. For model structure and component visualization, we leverage the implementation of TensorSpace. This visualization technique helps users work with the model graph in a more interactive way compared to node-link graphs proposed in TensorBoard. Intermediate outputs in each layer are rendered together with model components. Engineers can click on the model to expand the detailed components in each layer, drag and spin the model to check it from different angles, zoom in to look into a single layer, or zoom out to have a global view.

To make \textit{feature 2} more interactive, inspired by mutation analysis proposed in \cite{shen2018munn}, engineers can click on filters in a expanded layer to tell NeuralVis which filters should be frozen in order to implement model mutation in Figure \ref{fig:neuralvis}. Frozen filters are marked as grey indicating that it is inactive. As illustrated in \textit{Practice 2}, no activation status will be passed to the next layer by these frozen filters. Any changes brought to the following layers are visualized as a response to the user's action.

\subsection{The comparison panel}
Inspired by comparative research, software engineers can compare the output of multiple inputs with NeuralVis as stated in \textit{Practice 1}. By selecting a pair of inputs from the sample list, section C will pop out at the bottom of section B. Layers to intercept need to be selected. Activation status of that specified layer will be visualized. In this way, engineers can compare the intermediate output between the original input and a corresponding adversarial sample, similarities and differences observed will bring them intuition about hidden layers.

Since feature maps are of different sizes, it can be difficult for engineers to compare intermediate output in a relatively small panel. To make it more friendly, engineers are able to expand the comparison panel to full size.

\section{User Study}
To evaluate the usefulness and effectiveness of NeuralVis, we train LeNet~\cite{lecun2015lenet} on MNIST~\cite{deng2012mnist} and VGG-12~\cite{simonyan2014very} on CIFAR-10~\cite{krizhevsky2014cifar} to conducted a task-based user study with ten participants.
We used MNIST since it is easier for participants to draw handwriting images online. 
CIFAR-10 was used to illustrate the intuition brought to engineers in more complex situations. 


In this study, we design following tasks:
\begin{itemize}
\item {\textbf Task 1.} describing the structure, including the layers, specifications and their relationships based on both trained DNN models, 
\item {\textbf Task 2.} creating some inputs with the sketchpad and using the model to predict their label based on the trained LeNet;
\item {\textbf Task 3.} describing the differences between the intermediate outputs of two inputs based on the trained LeNet;
\item {\textbf Task 4.} highlighting  critical features in the original input that determine the prediction outputs based on the trained VGG-12;
\end{itemize}

We recruit ten master students of software engineering as participants. 
All participants are required to perform the designed tasks on the browser of their own laptop within one hour.

\begin{figure}[h]
  \centering
  \includegraphics[width=\linewidth]{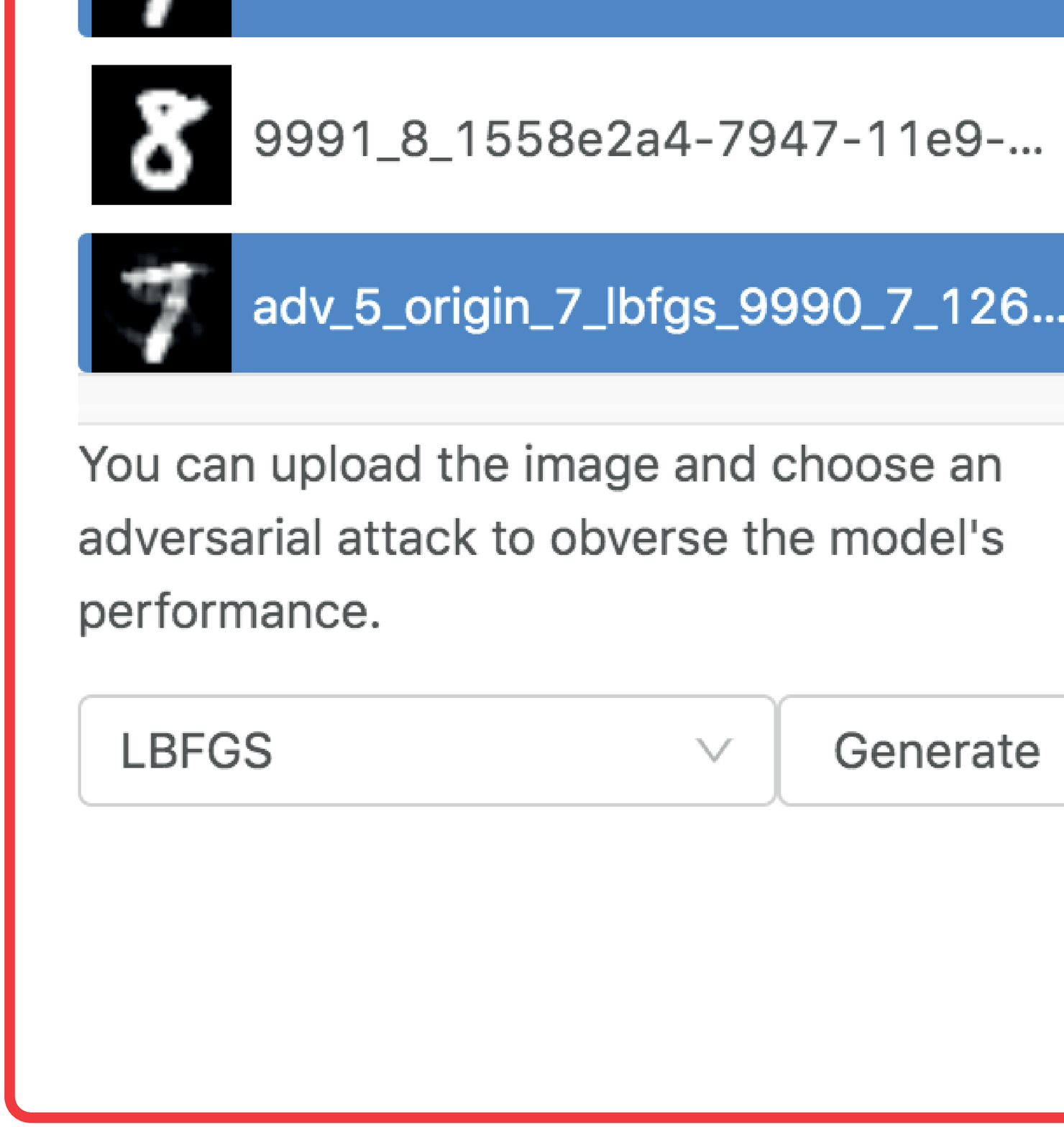}
  \caption{Visualization tasks performed in the study}
  \label{fig:userstudy}
\end{figure}

The study result shows that all developers can perform tasks in time with NeuralVis.
For task, as is shown in Figure \ref{fig:userstudy}, with the model file selected, structural information will be displayed. 
All participants can describe structure of used models.

All participants can  perform the tasks in time with NeuralVis. For task 1, as is shown in Figure \ref{fig:userstudy}, with the model file selected, structural information will be displayed. 
All participants were able to describe structure of used models. 
For task 2, participants manually write some digits on the sketchpad, and all of them successfully finish this task within several seconds.
For task 3, participants report that ``they can tell the feature that influences the final output".
Note that if a layer is frozen by users, no activation information is passed to the next layer. 
In this task, intermediate outputs are inspected, which offered the participant a chance to view the differences in hidden layers via comparative research. 
We analyze the result submitted by all participants and find that there were some filters participants considered to be important and filters that are useless according to the filters they highlighted, which proves that NeuralVis brings them intuition in understanding importance of neurons inside a neural network.




After all participants finish all tasks, we also interview them to collect feedback on NeuralVis. 
The feedback shows the comparison of intermediate output is critical for the analysis of the behavior of DNN models. 

\section{Conclusions}
In this paper, we present NeuralVis, a web-based tool for visualizing trained neural networks to enhance engineers' understanding on both model structure as well as data transformation.
Also, NeuralVis provides the functionality of comparing intermediate output between a pair of inputs. Based on this functionality, engineers can identify critical features that determine the prediction results. 
Usefulness and effectiveness of the tool are evaluated through a user study where engineers are able to understand the model structure and highlight important neurons when performing tasks.

Because NeuralVis is instance-based when visualizing behaviors, input selection counts a lot with respect to efficiency and efficacy. Similar to prioritize a test case during the test, we plan to guide engineers in choosing an input by labeling them with different priorities\cite{shi2019deepgini}.
\bibliographystyle{ACM-Reference-Format}
\bibliography{reference}

\end{document}